\documentclass[12pt]{article}
\usepackage{amsmath}
\usepackage{amssymb}
\usepackage{graphicx}
\usepackage{yf}

\newcommand{\guqq}{g_{\eta_{u} }}
\newcommand{\gsqq}{g_{\eta_{s} }}
\newcommand{\etau}{\eta_{u}}
\newcommand{\etas}{\eta_{s}}
\newcommand{\fu}{f_{u}}
\newcommand{\fs}{f_{s}}
\newcommand{\m}[1]{m_{#1}}
\newcommand{\diag}{{\rm diag}}
\newcommand{\comment}[1]{}
\newcommand{\Journal}[5]{#1~#2~\textbf{#3}, #5 (#4)}

\begin{document}
\title{Transition form factors and light cone distribution amplitudes
of pseudoscalar mesons in the chiral
quark model}
\author{A. E. Dorokhov, M.K. Volkov, V.L. Yudichev\\[5mm]
\itshape Joint Institute for Nuclear Research, Dubna, Russia \\
}
\date{}
\maketitle

\Abstract{
It is shown that a chiral quark model of the Nambu--Jona-Lasinio type
can be used to describe "soft"-momenta parts of the amplitudes with
large momentum transfer.
As a sample, the processes $\gamma^\ast\to\gamma (\pi,\eta,\eta')$ where
one of the photons, $\gamma^\ast$, has large space-like virtuality is investigated.
The $\gamma^\ast\to\gamma (\pi,\eta,\eta')$ transition form factors
are calculated for a wide region of transferred momenta.
The results are consistent with the calculations performed in
the instanton-induced chiral quark model
and agree with experimental data.
The distribution amplitudes of pseudoscalar mesons are derived.
}

\clearpage
\section{Introduction}
Effective chiral quark models (ECQM) are very useful tools for
the investigation of the nonperturbative sector of QCD.
Of particular interest are the Nambu--Jona-Lasinio (NJL) model and its extensions
\cite{Volk86a,Volk86,Ebert94,Vogl91,Kleva92}, where
the low-energy theorems are fulfilled, and
the mechanism of spontaneous breaking of chiral symmetry (SBCS)
is realized in a simple and transparent way.
The internal properties of the ground states of scalar,
pseudoscalar, vector, and axial-vector mesons
(such as masses, radii, polarizabilities, etc.) as well as
all the main decays and other low-energy strong and electroweak interactions of mesons are
satisfactorily described in NJL.
Recently, a noticeable progress has been achieved
also in constructing a $U(3)\times U(3)$ NJL model
for the description of first radial excitations of  scalar,
pseudoscalar, and vector mesons, including even the
lightest scalar glueball
\cite{Volk2000,Volk2001,Volk2001a}.

As a rule, the NJL model is used in the low-energy
physics, in particular, for the description of processes with
a low momentum transfer ($\leqslant 1$ GeV).  And, in view of the above-mentioned
success  of the NJL model, it is quite interesting to extend the
range of its application and to try, in particular, to
describe  some processes with a large momentum transfer.
Indeed, for many of such processes, the factorization theorem
\cite{Chernyak77,Efremov80,Brodsky80} can be
applied, which allows one to rewrite the amplitude of the
process as a convolution of "hard" and "soft" parts.
The "hard" part is described by the perturbative QCD (pQCD),
whereas the "soft" part requires a nonperturbative approach.
Usually, in QCD, the nonperturbative dynamics of quarks inside a meson is parameterized by
distribution amplitudes (DA) \cite{Chernyak77}, the exact form of which
cannot be derived from pQCD.
However, an approximation for DA can be obtained
in the QCD sum rules approach (see e.g.\cite{MikhRad90,MihRad92,RadRus96}).
On the other hand, one can calculate the "soft" part of the amplitude
within a chiral quark model,  where
the dynamics of (constituent) quarks inside a meson is described by the corresponding
quark-meson vertex instead of  DA
(however, one can  restore the shape of DA on the base of quark-model calculations).
The validity of this approach is investigated in the present work.

In our paper, we calculate the form factors that describe
the transition processes  $\gamma^\ast\to
P\gamma$ or $\gamma^\ast\gamma\to P$, where $P$ is a pseudoscalar meson,
for a wide range of transferred space-like momenta.
Of interest is  the kinematic region where one of the
photons is not on mass-shell and has a large space-like
virtuality.
It is shown also that our results do not contradict both the experimental data
and other theoretical models
. As to experiment, we refer to the data on the $\gamma^\ast\to\pi\gamma$,
$\gamma^\ast\to\eta\gamma$, and $\gamma^\ast\to\eta'\gamma$
transition processes reported by the CLEO collaboration
\cite{CLEO}. For comparison with other theoretical models, we choose the
results recently obtained in the framework of the instanton-induced chiral quark
model (IQM) \cite{DOAnTo00}.

The structure of our paper is as follows.
In Sect.~2, we introduce the $\gamma^\ast\to P\gamma$
transition form factor. In Sect.~3, the part of the NJL Lagrangian
describing the quark-meson interaction
as well as the Lagrangian derived in the instanton vacuum model
are introduced and the asymptotic
behaviour  of the $\gamma^\ast\to P\gamma$ transition form factor for
pseudoscalar mesons at high space-like virtuality
of one the photons is investigated. In particular, the shapes of
DA of $\pi, \eta$, and $\eta'$ mesons are found.
In the last section, we discuss the obtained results and compare them with
experimental data. There is also given  an outlook of further
possible applications of our approach.

\section{The $\gamma^\ast\to P\gamma$ transition form factor}
 The transition process
 $\gamma ^{\ast }(q_{1})\rightarrow P(p)\gamma ^{\ast }(q_{2}) $,
where the final state mesons with the momentum $p$ are, respectively,
 $P=\pi ^{0},\ \eta ,\ \eta ^{\prime }$,
and $q_{1}$ and $q_{2}$ are photon
momenta, is described by the amplitude
\begin{equation}
\mathcal{T}\left( \gamma ^{\ast }\left( q_{1},e_{1}\right) \rightarrow
 P\left( p\right) \gamma ^{\ast} \left(q_{2},e_{2}\right)  \right) =
{\mathcal{F}}_P(q_1^2,q_2^2,p^2)\epsilon _{\mu \nu \rho \sigma}e_{1}^{\mu}e_{2}^{\nu }
q_{1}^{\rho }q_{2}^{\sigma }
\label{amplitudePgg}
\end{equation}
where $e_{i}(i=1,2)$ are the photon polarization vectors; ${\mathcal{F}}_P(q_1^2,q_2^2,p^2)$
is the transition form factor; and $\epsilon _{\mu \nu \rho \sigma}$
is the fully anti-symmetric tensor.

Theoretically,
at zero virtualities, the form factor
\begin{equation}\label{pgg}
  {\mathcal{F}}_P(0,0,0)=\frac{1}{4\pi^2f_P}
\end{equation}
is related to the axial anomaly \cite{Adler,Jackiw}.
Here $f_P$ is a pseudoscalar meson weak decay constant defined
by the well known PCAC relation (for the pion,
$f_{\pi }=93$ {\rm MeV}). At asymptotically
large photon virtualities, its behaviour is predicted by
pQCD \cite{BrLep79}  and
depends crucially on the internal  meson dynamics
parameterized by a nonperturbative
DA, $\varphi _{P}^{A}(x)$, with $x$
being a fraction of the meson momentum, $p$, carried by a
quark.

Further, it is convenient to
parameterize the photon virtualities as
$q_{1}^{2}=-(1+\omega )Q^{2}/2$ and
$q_{2}^{2}=-(1-\omega )Q^{2}/2$, where $Q^2$
and $\omega$ are, respectively, the total virtuality of the photons and the
asymmetry in their distribution:
\begin{equation}
Q^{2}=-(q_{1}^{2}+q_{2}^{2})\geqslant 0, \;\;\;\;{\rm and}\;\;\;\; \omega
=(q_{1}^{2}-q_{2}^{2})/(q_{1}^{2}+q_{2}^{2}),\quad |\omega|\leqslant 1  \label{Omega}
\end{equation}

Recent analysis of the experimental data on the form
factors ${\mathcal{F}}_P$ for small virtuality of one of the photons,
$q_{2}^{2}\approx 0$, with the virtuality of the other photon
being scanned up to $8$ GeV$^{2}$ for the pion, 22 GeV$^{2}$ for the $\eta$,
and  30 GeV$^{2}$ for the $\eta'$ mesons, respectively,
has been published  by the CLEO collaboration \cite{CLEO}.
According to this analysis, the process
$\gamma^{\ast}\rightarrow P\gamma$ $\left( |\omega |=1\right) $ can be
fitted by a monopole form factor:
\begin{eqnarray}
&&\left. {\mathcal{F}}_P(q_{1}^{2}=-Q^{2},q_{2}^{2}\approx0,p^2)\right| _{\rm fit}= \frac{
g_{P\gamma \gamma }}{1+Q^{2}/\mu_{P }^{2}}, \\
&&g_{\pi\gamma\gamma}\simeq0.27\ \mbox{GeV}^{-1},
\quad g_{\eta\gamma\gamma}\simeq0.26\ \mbox{GeV}^{-1},
\quad g_{\eta'\gamma\gamma}\simeq0.34\ \mbox{GeV}^{-1},
 \nonumber\\
&&
\mu_{\pi}\simeq 0.78\ \mbox{GeV},\quad
\mu_{\eta}\simeq 0.77\ \mbox{GeV},\quad
\mu_{\eta'}\simeq 0.86\ \mbox{GeV},\nonumber
\label{Fpiggfit}
\end{eqnarray}
where $g_{P \gamma \gamma }$ are the two-photon meson decay constants.

In the lowest order of pQCD,  the light-cone Operator Product Expansion
(OPE) predicts the high $Q^{2}$ behaviour of the form factor
as follows \cite{BrLep79}:
\begin{equation}
\left. {\mathcal{F}}_P(q_{1}^{2},q_{2}^{2},p^2=0)\right| _{Q^{2}\rightarrow \infty}
=J_P\left( \omega \right) \frac{f_{P}}{Q^{2}}
+O\left(\frac{\alpha_s}{\pi}\right)
+O\left(\frac{1}{Q^4}\right),  \label{AmplAsympt}
\end{equation}
with the asymptotic coefficient given by
\begin{equation}
J_P\left( \omega \right) =\frac{4}{3} \int\limits_{0}^{1} \frac{dx}{1-\omega^2 (
2x-1)^2}\varphi_P^{A}(x),  \label{J}
\end{equation}
where $\varphi_P^{A}(x)$ is the leading-twist meson light-cone DA
 normalized by $\int\limits_{0}^{1}
dx\,\varphi_P^{A}(x)=1$.

In (\ref{J}), the asymptotic coefficient is expressed in terms of
DA. Alternatively, as it was mentioned in the Introduction,
$J_P(\omega)$ can be calculated directly either from the NJL model or from
IQM. In Sect.~3, one will see that, in both models,  $J_P(\omega)$
can be rewritten in the form (\ref{J}), and thus the shape
of DA is extracted.

Since the meson DA reflects the internal
nonperturbative meson dynamics, the prediction of the value of
$J_P\left(\omega \right)$ is rather a nontrivial task, and its accurate measurement
would provide quite valuable information.
It is important to note that, for the considered transition process, the
leading asymptotic term of pQCD expansion (\ref{AmplAsympt}) is not
suppressed by the strong coupling constant $\alpha_s$. Hence, the pQCD
prediction (\ref{AmplAsympt}) can become reasonable
at the highest of the presently accessible momenta
$Q^{2}\sim 10$ GeV$^2$.
At asymptotically high $Q^{2}$, the DA evolves to
$\varphi_P^{A,\rm asympt}(x)$
$=6x(1-x)$ and $J_P^{\rm asympt}\left( \left| \omega \right|=1\right) =2$.
The fit of CLEO data for the pion  corresponds to
$J_\pi^{\rm CLEO}\left( \left| \omega \right| \approx 1\right) = 1.6\pm 0.3$,
indicating that already at moderately high
momenta this value is not too far from its
asymptotic limit.

However, since the pQCD evolution of DA reaches the asymptotic regime
very slowly, its exact form at moderately high $Q^2$ does not coincide with
$\varphi_P^{A,\rm asympt}(x)$. At lower $Q^{2}$, the power corrections to the
form factor become important. Thus, the study of the behaviour of the transition
form factor at all experimentally accessible $Q^{2}$ is the subject of
nonperturbative dynamics.
So, the theoretical determination of the transition form factor is still
challenging, and it is desirable to perform direct calculations of
${\mathcal{F}}_P(q_{1}^{2},q_{2}^{2},p^2) $ without  {\it a priori} assumptions about
the shape of the meson DA.

The asymptotic coefficient of
the light-cone transition form factor in the symmetric kinematics, $q_{1}^{2}=q_{2}^{2}$%
, $(\omega =0)$\ at high virtualities, is given by the integral defining
$f_P$.
In the other extreme limit, where one
photon is  real $(|\omega| =1)$, the asymptotic coefficient is
proportional to
$\int\limits_{0}^{1}\frac{dx}{x}\,\varphi _{P}^{A}(x)$
and thus is very sensitive to a detailed form of the DA.

In ref.~\cite{MikhRad90}, some progress was achieved by using a
refined technique based on the OPE with nonlocal condensates
\cite{MihRad92}, which is equivalent to the inclusion of the whole
series of power corrections. By means of the QCD sum rules with
nonlocal condensates, it was shown that this approach works in
almost the whole kinematic region $\left| \omega \right| \lesssim 1$,
and that for high values of the asymmetry parameter $\left| \omega
\right| \gtrsim 0.8$, the pion transition form factor
is very sensitive to the nonlocal structure of the QCD vacuum.

\section{The $\gamma^\ast\to P\gamma$ transition form factor in ECQM models}

\subsection{The NJL model Lagrangian}
Let us consider the piece of the effective quark-meson Lagrangian
that is necessary in our calculations. It has the following form:
\begin{eqnarray}
L&=&L_0+L_{\rm int}\label{S_Q_P_begin}\\
L_0&=&\bar{q}(i\!\not\!\partial-m)q,\label{S_Q_P0}\\
L_{\rm int}&=&L_1+L_2+L_3+\delta L,\label{S_Q_Pint}\\
L_1&=&\bar{q}i\gamma_5\left(
g_{\pi}\sum_{a=1}^{3}\lambda^{a}\pi ^{a}+
\guqq \lambda_{\rm u}\etau+
\gsqq \lambda_{\rm s}\etas\right)q,\label{S_Q_PL1}\\
L_2&=&\bar{q}i\!\not\!\partial\gamma_5\left(
f_\pi^{-1}(1-Z_u^{-1})\sum_{a=1}^{3}\lambda^{a}\pi ^{a}+
 f_u^{-1}(1-Z_u^{-1})\lambda_{\rm u}\etau+
f_s^{-1}(1-Z_s^{-1}) \lambda_{\rm s}\etas\right)q,\label{S_Q_PL2}\\
L_3&=&\bar{q}\hat{\mathcal{Q}}\!\not\!\!\!\mathcal{A} q,
\label{S_Q_P_end}
\end{eqnarray}
where $m$ is the diagonal $3\times 3$-flavor matrix of constituent quark masses,
$m=\diag(\m{u},\m{d},\m{s})$, (we consider the case of approximate
isotopic symmetry $\m{u}=\m{d}$);
$\lambda_a$ are the Gell-Mann matrices, $\lambda_{u}=
(\sqrt{2}\lambda_0+\lambda_8)/\sqrt{3}$, $\lambda_{s}=
(-\lambda_{0}+\sqrt{2}\lambda_8)/\sqrt{3}$;
 $q$ and $\pi$ are, respectively, the quark and pion,  and
$\etau$ and $\etas$ are pure $\bar uu$ and $\bar ss$ pseudoscalar meson states.

The fields $\eta_{\rm u}$ and $\eta_{\rm s}$ in (\ref{S_Q_PL1}) and (\ref{S_Q_PL2})
are not  physical, because
they are subject to singlet-octet mixing.
Here, it is assumed that the terms responsible for the singlet-octet mixing
are accumulated in the term%
\footnote{ The singlet-octet mixing appears once
the pseudoscalar gluon anomaly is taken into account
\cite{Volk86}. One can also obtain the mixing after introducing
the 't Hooft term into the quark Lagrangian \cite{NuovoCim}.
} $\delta L$ (see (\ref{S_Q_Pint})), the account
of which results in the following relation between "nonphysical"  $\eta_u$, $\eta_s$
and  "physical"  $\eta$, $\eta'$ meson fields:
\begin{eqnarray}
\etas&=&\eta\cos(\theta_0-\theta)  -\eta'\sin(\theta_0-\theta), \\
\etau&=&\eta\sin(\theta_0-\theta)  +\eta'\cos(\theta_0-\theta),
\label{somixing}
\end{eqnarray}
where  $\theta=-19^\circ$ is the singlet-octet mixing angle,
and $\theta_0\approx 35.5^\circ$ is the ideal mixing
angle \cite{Volk86,NuovoCim}.

The quark-meson coupling constants (see \cite{Volk86})
are defined as follows:
\begin{equation}\label{gus}
g_{\pi}=g_u\sqrt{Z_\pi},\quad g_{\etau }= g_{\rm u}\sqrt{Z_{\eta_u}},\quad
g_{\etas }= g_{\rm s}\sqrt{Z_{\eta_s}},
\end{equation}
where we introduced $g_{\rm u}$ and $g_{\rm s}$:
\begin{equation}
g_a^{-2}=\frac{4 N_{\rm c}}{(2\pi)^4}\int
\frac{\theta(\Lambda_{\rm NJL}^2-k^2)}{(k^2+m_a^2)^2} d_e^4 k,\quad (a={\rm u,s}).
\label{gq}
\end{equation}
The integration is performed in the Euclidean metric.
The divergence is eliminated by a simple $O(4)$-symmetric
cut-off on the scale $\Lambda_{\rm NJL}$ that characterizes the domain of SBCS.

The terms with derivatives of meson fields in $L_2$ (see (\ref{S_Q_PL2})) appear
because of $\pi-a_1$ transitions \cite{Volk86},
which also results in  additional renormalization factors $Z_a$ in
$g_\pi, \guqq, \gsqq$. Further, we assume that $Z_a$
for different mesons are approximately equal (see (\ref{gus})):
\begin{equation}\label{Z}
 Z_{\eta_s}\approx Z_{\eta_u}\approx Z_\pi\equiv Z=
\left(1-\frac{6m_u^2}{M_{a_1}^2}\right)^{-1}\approx 1.45.
\end{equation}
Here, $M_{a_1}=1.23$ GeV is the mass of the $a_1$-meson \cite{PDG}.

The term $L_3$ in (\ref{S_Q_P_end}) describes the electromagnetic interaction
of quarks. The photon fields are denoted by $\mathcal{A}$,
and $\hat{\mathcal{Q}}$ stands for the charge matrix:
\begin{equation}
\hat{\mathcal{Q}}=\frac{e}{2}\left(\lambda_3+\frac{\lambda_8}{\sqrt{3}}\right),
\end{equation}
where $e$ is the elementary electric charge ($e^2=4\pi \alpha$ where $\alpha^{-1}\approx 137$).

The values of $\Lambda_{\rm NJL}$ and $\m{u}$ are
fixed by two equations \cite{Volk86,NuovoCim}:
i) the Goldberger-Treiman relation $\m{u}=g_{\pi}f_\pi$
and
ii) the $\rho$-meson decay constant \cite{Volk86,Eguchi,Kikkawa}
\begin{equation}\label{grho}
g_\rho=\sqrt{6}g_u
\end{equation}
  whose value
$6.1$ is well known from the experimentally observed decay $\rho\to\pi\pi$.
Taking into account these equations and the expression for $Z$ (see (\ref{Z})),
one finds the constituent $u$-quark mass:
\begin{equation}\label{mumass}
  m_u^2=\frac{M_{a_1}^2}{12}\left(1-\sqrt{1-\frac{4g_\rho^2 f_\pi^2}{M_{a_1}^2}}\right)
\end{equation}
with the value $\m{u}=280$ MeV.
Equating the left-hand side of (\ref{grho}) to its experimental value and
using the definition of $g_u$ (see (\ref{gus})) with $m_u=280$ MeV,
one obtains: $\Lambda_{\rm NJL}=1.25$ GeV \cite{Volk86}.
The mass of the strange quark is fixed by the kaon mass%
\footnote{The strange quark mass can be fixed also from the
$\phi$-meson mass \cite{VolkovIvanov}},
$\m{s}=425$ MeV \cite{NuovoCim}.

\subsection{Meson transition $\protect\gamma ^{\ast }
\rightarrow \protect P\protect\gamma ^{\ast}$ form factor}

\begin{figure}[h]
\caption{}
\begin{center}
\includegraphics[angle=90,scale=0.75]{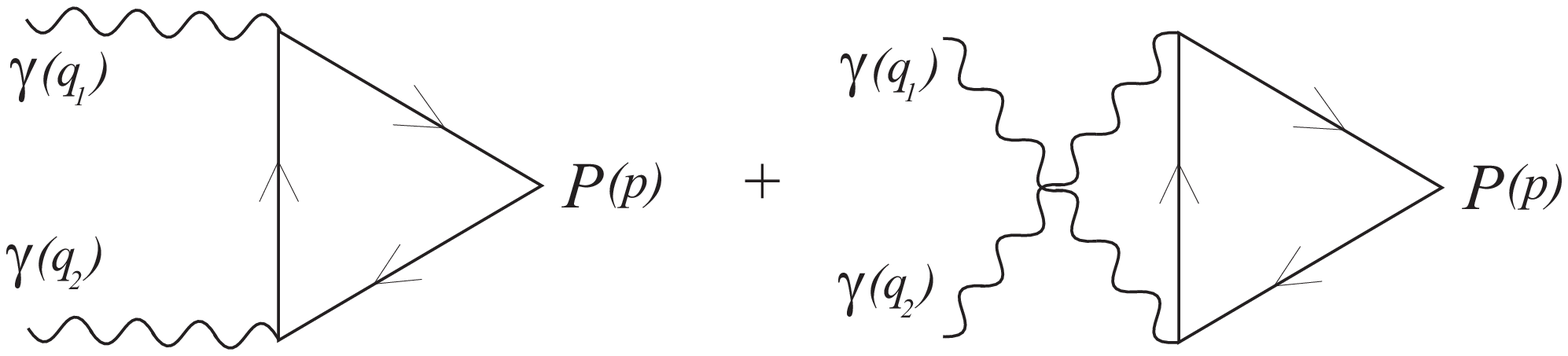}
\end{center}
\label{Pgg}
\end{figure}

Let us consider  the $\gamma ^{\ast }\to P\gamma ^{\ast }$
invariant amplitude corresponding to the triangle diagrams
shown in Fig.~\ref{Pgg}:
\begin{equation}
\mathcal{T}\left( \gamma ^{\ast }\left( q_{1},e_{1}\right)
\rightarrow P\left( p\right)
\gamma ^{\ast} \left(q_{2},e_{2}\right)\right) =t_{P
\gamma \gamma }(q_{1},e_{1};q_{2},e_{2})+ t_{P \gamma \gamma}(q_{2},e_{2};q_{1},e_{1}),
\label{amplitude}
\end{equation}
\begin{eqnarray}\label{amplPgg}
&&t_{P \gamma \gamma }(q_{1},e_{1};q_{2},e_{2})=
-N_{c} g_{P}\mathcal{Q}_P\times \nonumber\\
&&\quad\times \int \frac{d^{4}k}{(2\pi )^{4}}
\mbox{tr}\{i\gamma_{5} S(k+p/2;m_a){\hat{e}}_{1}S(k-(q_{1}-q_{2})/2;m_a)
{\hat{ e}}_{2}S(k-p/2;m_a)\}
\end{eqnarray}
where $\mathcal{Q}_P$ depends on the electric charges and flavors  of quarks that constitute the meson:
 $\mathcal{Q}_{\pi^0}=1/3$ for $\pi^0$, $\mathcal{Q}_{\eta_u}=5/9$ for $\eta_u$,
and $\mathcal{Q}_{\eta_s}=-\sqrt{2}/9$ for $\eta_s$;  $S$ is the quark propagator:
\begin{equation}
S^{-1}(k;m_a)=\not\!k-m_a
\end{equation}
with the constituent quark mass $m_{a}=\m{u}$ for $P=\pi$ or $P=\etau$ and $m_{a}=\m{s}$
for $P=\etas$.
Comparing (\ref{amplPgg}) with (\ref{amplitudePgg}), one obtains:
\begin{equation}
{\mathcal{F}}_P(q_{1}^{2},q_{2}^{2},p^2)=\frac{g_{P}}{2\pi ^{2}}m_{a}I_{P
\gamma \gamma }(q_{1}^{2},q_{2}^{2},p^{2}).  \label{invAmpl}
\end{equation}

The Feynman integral $I_{P \gamma \gamma}
(q_{1}^{2},q_{2}^{2},p^{2}) $ is given (in Euclidean metric) by
\begin{eqnarray}
&&\kern-3mm I_{P \gamma \gamma }(q_{1}^{2},q_{2}^{2},p^{2})=\label{Ipigg}\\
&&\;=\int \frac{d^{4}k}{\pi^{2}}
\frac{\theta(\Lambda_{\rm NJL}^2-k^2)}{[m_{a}^{2}+(k+p/2)^{2}]
[m_{a}^{2}+(k-p/2)^{2}]
[m_{a}^{2}+(k-(q_{1}-q_{2})/2)^{2}]}. \nonumber
\end{eqnarray}

In the chiral limit $p^2=0$, when both photons are on-mass-shell
($q_1^2=q_2^2=0$),  integral (\ref{Ipigg}) becomes very simple.
Formally, it is finite, and one can set the UV cut-off $\Lambda_{\rm NJL}$ to infinity,
and thus obtain that it is equal to $1/(2m_a^2)$. As a result,
one reproduces the well-known result for the decay $\pi^0\to\gamma\gamma$
(see (\ref{pgg})).
For the $\eta$ and $\eta'$ mesons, the result is similar,
and the only difference  is that  the singlet-octet mixing
should be taken into account:
\begin{equation}
{\mathcal{F}}_\eta(0,0,0)=\frac{1}{4\pi^2 \tilde f_\eta},\quad
{\mathcal{F}}_{\eta'}(0,0,0)=\frac{1}{4\pi^2 \tilde f_{\eta'}},
\end{equation}
\begin{eqnarray}
\tilde f_\eta^{-1} &=& \frac{5}{3f_{u}}\sin(\theta_0-\theta)-
\frac{\sqrt{2}}{3f_{s}}\cos(\theta_0-\theta)\label{f0eta}\\
\tilde f_{\eta'}^{-1}&=&\frac{5}{3f_{ u}}\cos(\theta_0-\theta)+
\frac{\sqrt{2}}{3f_{s}}\sin(\theta_0-\theta)\label{f0etaprime}.
\end{eqnarray}
Here, the meson weak decay constants are
$
 \fu\equiv f_{\pi}, \quad\mbox{and}\quad \fs=m_s/g_s\approx 1.25 \fu.
$
Thus, we have $\tilde f_\eta=83$ MeV and $\tilde f_{\eta'}=73$ MeV.
(For the discussion of singlet-octet mixing see \cite{Feldmann})

Let us now consider the high virtuality region: $Q^2\to\infty$ (see (\ref{Omega})
for definition) .
We estimate the asymptotics of the transition form factor.
Let us rewrite the expression for integral (\ref{Ipigg})
by using the
Feynman $\alpha -$parameterization for the denominators and integrating over
the angular variables. Then, the corresponding integral $I_{P \gamma
\gamma }$ is given by
\begin{eqnarray}
&&I_{P \gamma \gamma }(q_{1}^{2},q_{2}^{2},p^{2})=\int\limits_{0}^{\Lambda_{\rm NJL}^2/m_q^2 }\frac{
u\, du}{m_{q}^{2}+u-\frac{p^{2}}{4}}\times\nonumber\\
&&\times\int\limits_{0}^{1}d\alpha \left[ \frac{1}{\sqrt{b^{4}-a_{+}^{4}}\left( b^{2}+\sqrt{
b^{4}-a_{+}^{4}} \right) }+\frac{1}{\sqrt{b^{4}-a_{-}^{4}}\left( b^{2}+\sqrt{
b^{4}-a_{-}^{4}} \right) }\right] ,  \label{Igpigg3}
\end{eqnarray}
where $u=k^2$ and
\begin{equation}
b^{2}=m_{a}^{2}+u+\frac{1}{2}\alpha Q^{2}-\frac{1}{4}\left( 1-2\alpha
\right) p^{2},\quad a_{\pm }^{4}=2u\alpha Q^{2}\left( \alpha \pm \omega
\left( 1-\alpha \right) \right) -\left( 1-2\alpha \right) up^{2}.
\label{notas2}
\end{equation}

In this way, the expression (\ref{Igpigg3}) can be safely analyzed in the
asymptotic limit of high total virtuality of the photons $Q^{2}\rightarrow
\infty $. Moreover, the integral over $\alpha $ can be taken analytically,
leading, in the chiral limit $p^2=0$, to the asymptotic expression
given by (\ref{AmplAsympt}), where (see \cite{DOAnTo00})
\begin{eqnarray}
&&J_P(\omega)\equiv J_{P,\rm np}\left( \omega \right) =\frac{g_P^2 \mathcal{Q}_P Z}{8\pi^2\omega }
\left\{ {\ \int\limits_{0}^{\Lambda_{\rm NJL}^2/m_q^2} \frac{du}{1+u} \ln %
\left[ \frac{1+u\left( 1+\omega \right)}{1+u\left( 1-\omega \right) } \right]
}\right\}  \label{Igpigg2} \\
&& a={\rm u} \;\mbox{if}\; P=\pi,\etau,\; \mbox{and}\; \quad a={\rm s}
\;\mbox{if}\; P=\etas.
\end{eqnarray}
The $\eta$ and $\eta'$ mesons appear as mixed $\etau$ and $\etas$ states
(see (\ref{somixing})), and for them, one has:
\begin{eqnarray}
&&J_\eta(\omega)=c_1\;J_{\etau}(\omega)+c_2
\; J_{\etas}(\omega),\\
&&J_{\eta'}(\omega)=c_3\; J_{\etau}(\omega)+
c_4\; J_{\etas}(\omega),
\end{eqnarray}
where the coefficients $c_i$ are:
\begin{equation}
\begin{array}{ll}
\displaystyle c_1=\frac{5f_{\rm u}}{3f_{\eta}}\sin(\theta_0-\theta),&
\displaystyle c_2=-\frac{\sqrt{2}f_{\rm s}}{3f_\eta}\cos(\theta_0-\theta),\\
\displaystyle c_3=\frac{5f_{\rm u}}{3f_{\eta'}}\cos(\theta_0-\theta),&
\displaystyle c_4=\frac{\sqrt{2}f_{\rm s}}{3f_{\eta'}}\sin(\theta_0-\theta),
\end{array}
\label{c}
\end{equation}
and the constants $f_{\eta}$ and $f_{\eta'}$ are defined as
\footnote{
Note that the definition of $f_{\eta}$ and $f_{\eta'}$
differs from that of $\tilde f_\eta$ and $\tilde f_{\eta'}$
(see (\ref{f0eta}) and (\ref{f0etaprime})).
}
\begin{eqnarray}
f_{\eta}&=& \frac{5}{3}f_{\rm u}\sin(\theta_0-\theta)-
\frac{\sqrt{2}}{3}f_{\rm s}\cos(\theta_0-\theta)=95\;\text{MeV},\label{feta}\\
f_{\eta'}&=&\frac{5}{3}f_{\rm u}\cos(\theta_0-\theta)+
\frac{\sqrt{2}}{3}f_{\rm s}\sin(\theta_0-\theta)=135\;\text{MeV}\label{fetaprime}.
\end{eqnarray}

One should also note
an extra factor $Z$ in the expression of
$J_P(\omega)$ in (\ref{Igpigg2}).
Analogously, the factor $Z$ appears in the amplitude of the decay $\pi\to\mu\tilde\nu_\mu$
that determines the pion weak coupling constant $f_\pi$.
To obtain a correct result, one should take account of $\pi-a_1$ transitions
by considering additional contributions going from the diagrams with axial-vector type vertices
(see the term $L_2$ in (\ref{S_Q_PL2})).
When calculating both the $\pi\to\mu\tilde\nu_\mu$ amplitude and
the $\gamma^\ast\to P\gamma^\ast$ transition form factor,
the account of $\pi-a_1$ transitions leads to
cancelation of the factor $Z$.

\subsection{Distribution amplitudes of pseudoscalar mesons. }

The quark-meson interaction in the NJL model \cite{Volk86}
is described by the vertices given in  (\ref{S_Q_P_begin})--(\ref{S_Q_P_end}).
One just has to calculate the integral (\ref{Ipigg}).
Formally, the integral (\ref{Ipigg})
is convergent, and an UV cut-off is not necessary here,
however, the extracting of its asymptotic
behavior at large $Q^2$ would
lead to logarithmic dependence $\sim (\ln Q^2)^2/Q^2$, which is not expected,
as it is known from QCD. Therefore, the UV cut-off in the NJL model should be
treated not only as a trick to make the integrals convergent,
but also as a way to take into account
the nontrivial nonlocal vacuum structure.
The UV cut-off also forbids the large momenta flow through  meson vertices.

From (\ref{Igpigg2}) it is clear that the prediction of the
nonperturbative approach to the asymptotic coefficient $J_P(\omega)$ is rather
sensitive to the ratio of the ultra-violet cut-off $\Lambda_{\rm NJL}$  to the value of the constituent mass $m_{a}$
 and to the relative distribution of the total virtuality among photons,
$\omega $. In particular, for the off-shell process $\gamma ^{\ast }\rightarrow \pi ^{0}\gamma
^{\ast }$ in the kinematic case of symmetric
distribution of photon virtualities, $q_{1}^{2}=q_{2}^{2}\rightarrow -\infty
$ ($\omega \rightarrow 0$), the result $J\left( \left| \omega \right| =0\right) =4/3$
obtained from (\ref{Igpigg2}) is in agreement with the OPE
prediction.

Integral (\ref{Ipigg}) is similar in its structure to the
integral arising in the lowest order of pQCD treating the quark-photon
interaction perturbatively. In the latter case, its asymptotic behavior is
due to the subprocess $\gamma^\ast(q_1) + \gamma^\ast(q_2)
\to \bar q(\bar xp) + q(xp) $
with $x$ ($\bar x=1-x$) being the fraction of the meson momentum $p$
carried by the quark produced at the $q_1$ ($q_2)$ photon vertex. The
relevant diagram is similar to the handbag diagram for hard exclusive
processes, with the main difference that one should use, as a
nonperturbative input, a quark-meson vertex instead of the meson DA.
As we see below, this similarity allows one to reconstruct
the  shape of the meson DA.

Both the expressions for $J_P$ derived within the  quark-meson
model (\ref{Igpigg2}) and from the light-cone
OPE (\ref{J}) can be put into the
common form
\begin{equation}
J_P\left( \omega \right) =\frac{2 }{3\omega }\int\limits_{0}^{1}\! d\xi R_P(\xi )\ln %
\left[ \frac{1+\xi \omega }{1-\xi \omega }\right]  \label{R}
\end{equation}
with
\begin{eqnarray}
&&\!\!R_{P}^{\rm pQCD}(\xi )=-\frac{d}{d\xi }\varphi _{P}^{A}\left( \frac{1+\xi }{2}
\right) \;\;{\rm and}\;\;R_{P,\rm np}(\xi )=16\pi^2 g_a^2\theta\left(
1-\frac{\xi }{1-\xi }\frac{m_a^2}{\Lambda_{\rm NJL}^2}\right) \frac{1}{1-\xi },\label{WF_VFdif}\\
&&\!\!\mbox{where}
\;\;\;0\leqslant \xi \equiv (2x-1)\leqslant 1 \nonumber
\end{eqnarray}
and similar expressions for $-1\leqslant \xi \leqslant 0$.
Equating both the
contributions, we find the meson DA
\begin{equation}
\varphi _{P }^{A}(x)= \frac{g_{a }^2}{4\pi^2}\int\limits_{|2x-1|}^{1}\!\!\theta\left(
1-\frac{y }{1-y }\frac{m_a^2}{\Lambda_{\rm NJL}^2}\right)\frac{dy}{1-y}.  \label{WF_VF}
\end{equation}
Thus, we show that  (\ref{Igpigg2}) obtained within the
NJL model is equivalent to the standard lowest-order pQCD result (\ref
{J}), with the only difference that the nonperturbative information
accumulated in the meson DA $\varphi _{P}^{A}(x)$ in pQCD  is represented in NJL by the
quark-meson vertex and is connected with the regularization procedure.

\begin{figure}
\caption{}
\includegraphics[angle=90,scale=0.7]{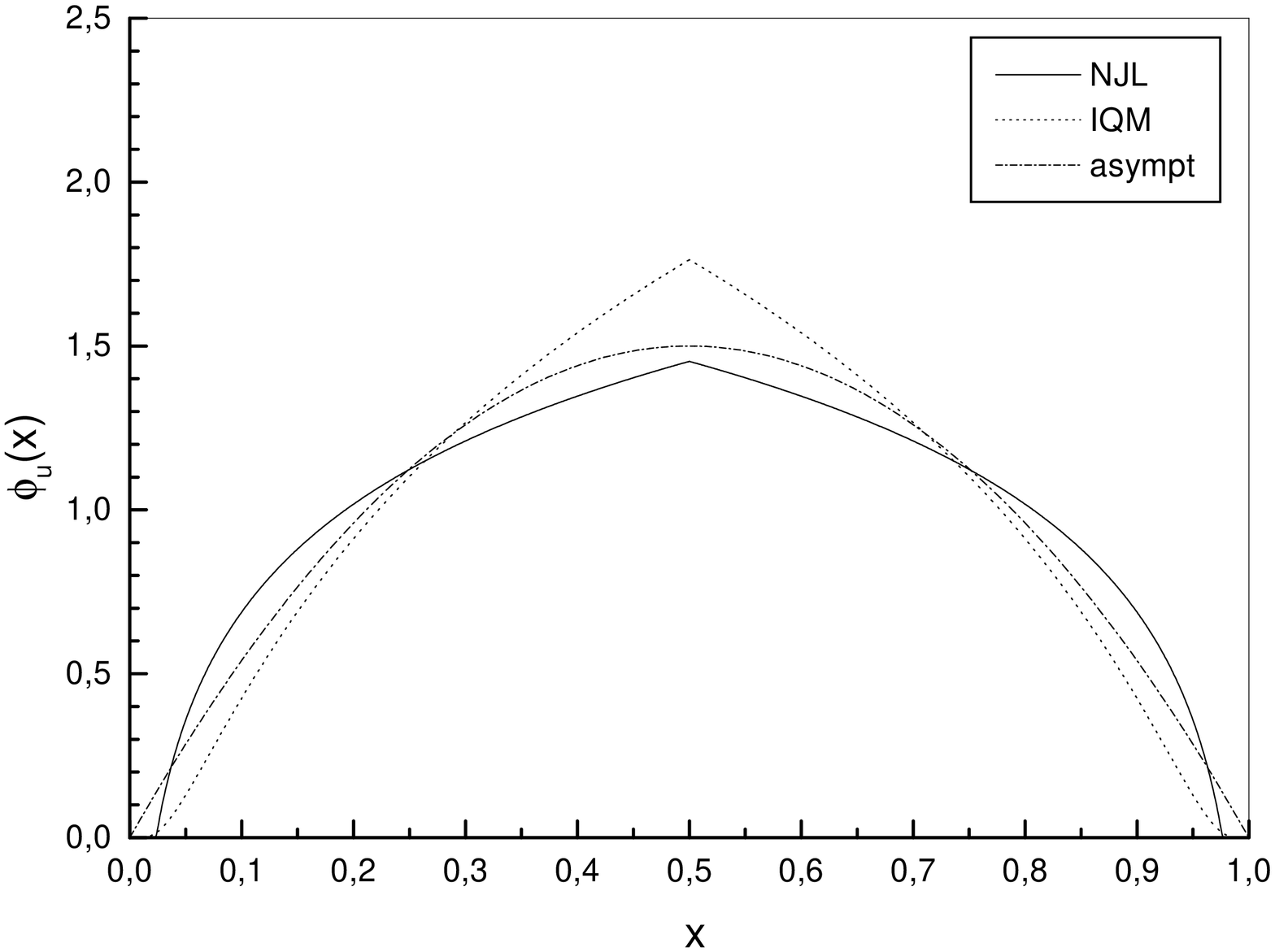}
\label{fiuplot}
\end{figure}

\begin{figure}
\caption{}
\includegraphics[angle=90,scale=0.7]{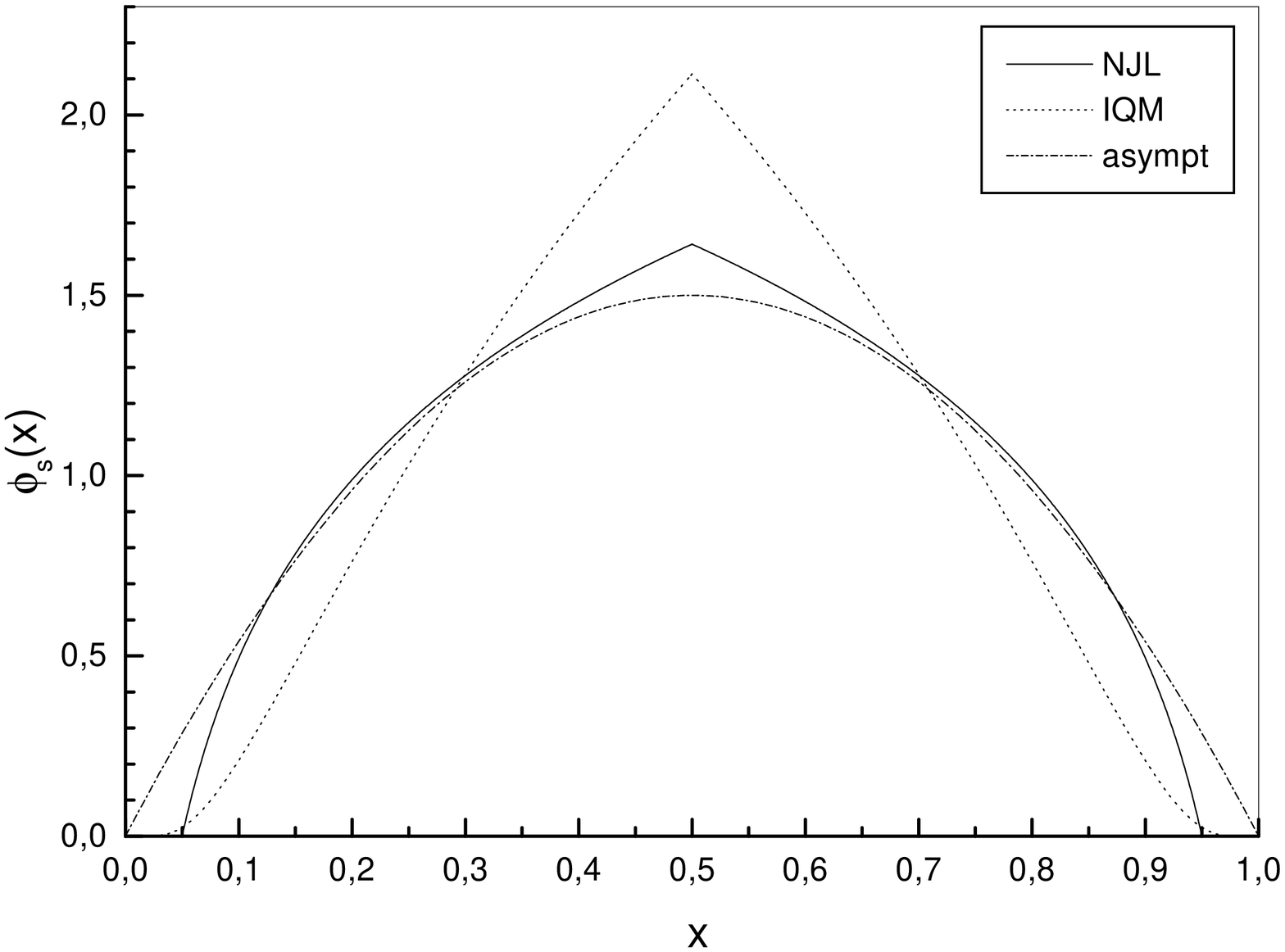}
\label{fisplot}
\end{figure}

\begin{figure}
\caption{}
\includegraphics[angle=90,scale=0.7]{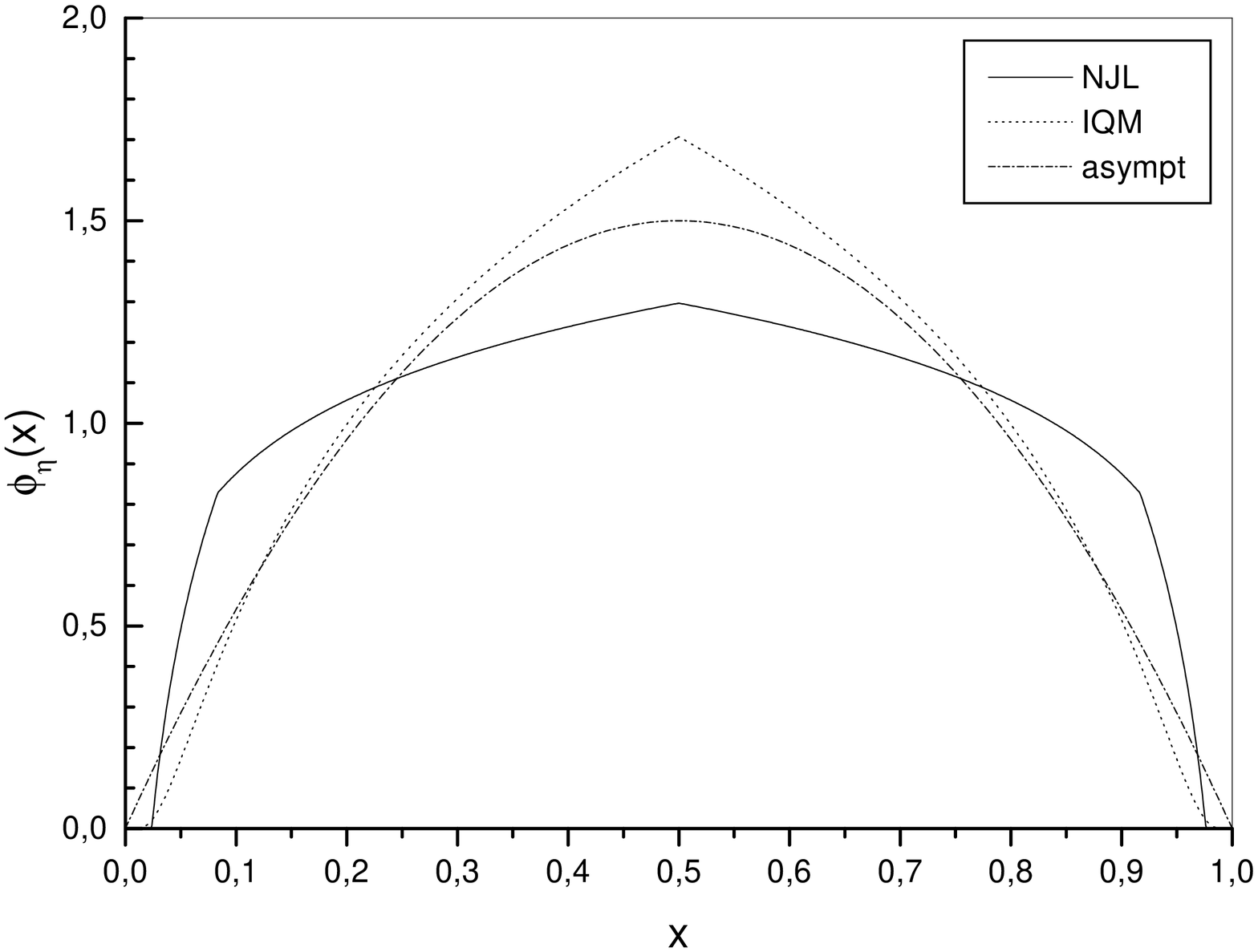}
\label{fietaplot}
\end{figure}

\begin{figure}
\caption{}
\includegraphics[angle=90,scale=0.7]{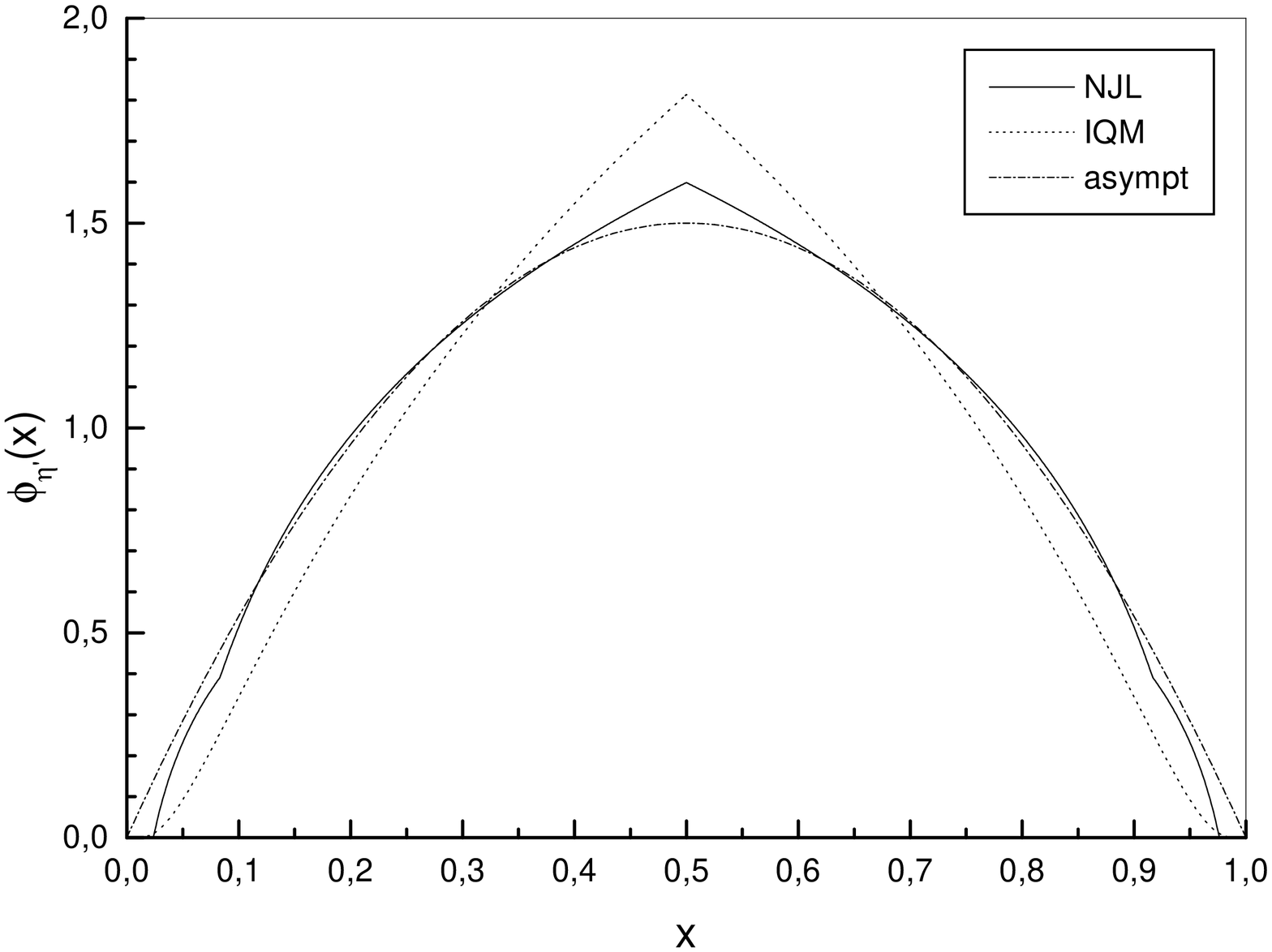}
\label{fietaprimeplot}
\end{figure}

Within the NJL model, from  (\ref{WF_VF}) one can easily obtain
 analytic expressions
for two special DA describing the distribution of
$u(d)$- and $s$-quarks, respectively,
\begin{equation}
\varphi_{ u}^A(x)=
\left\{
\begin{array}{ll}
\frac{\displaystyle\ln\left|\frac{1-\xi_{u}}{1-|2x-1|}
\right|}{\displaystyle\xi_{u}+\ln|1-\xi_{u}|},&
\quad |2x-1|\leqslant\xi_{u}\\[3mm]
0, &\quad |2x-1|>\xi_u
\end{array}\right.
\label{fiu}
\end{equation}
where the constant $\xi_{u}$ is defined as follows:
\begin{equation}
\xi_u=\frac{\Lambda_{\rm NJL}^2}{\Lambda_{\rm NJL}^2+\m{u}^2};
\end{equation}
\begin{equation}
\varphi_{s}^A(x)=
\left\{
\begin{array}{ll}
\frac{\displaystyle\ln\left|\frac{1-\xi_{s}}{1-|2x-1|}
\right|}{\displaystyle\xi_{s}+\ln|1-\xi_{s}|},&
\quad |2x-1|\leqslant\xi_{s}\\[3mm]
0, &\quad |2x-1|>\xi_{s}
\end{array}\right.,
\end{equation}
\begin{equation}
\xi_s=\frac{\Lambda_{\rm NJL}^2}{\Lambda_{\rm NJL}^2+\m{u}^2}.
\label{fis}
\end{equation}
For the pion, $\eta$, and $\eta'$ mesons, one has:
\begin{eqnarray}
\varphi_{\pi}(x)&=&\varphi_{u}^A(x),\\
\varphi_{\eta}(x)&=&c_1\;\varphi_{u}^A(x)+
c_2\varphi_{s}^A(x),\\
\varphi_{\eta'}(x)&=&c_3\;\varphi_{u}^A(x)+
c_4\varphi_{s}^A(x),
\end{eqnarray}
where the coefficients $c_i$ are defined in (\ref{c}).
The DA calculated in the NJL are shown in
Figs.~\ref{fiuplot}--\ref{fietaprimeplot}.

\subsection{Instanton-induced effective quark-meson Lagrangian}
Let us consider now the piece of the effective quark-meson Lagrangian
that appears in IQM.
The effective quark-meson dynamics can be summarized in the covariant
nonlocal action given by
\begin{eqnarray}
&&S_{\rm int}=
-\int\! d^4x\, d^4y\ F\left[x+y/2,x-y/2;\Lambda_{\rm IQM} ^{-2}\right]\times\nonumber\\
&&\quad\times\bar{q}
(x+y/2)\ i\gamma_5[
g_{\pi \bar{q}q}\sum_{a=1}^{3}\lambda^{a}\pi ^{a}(x)+
\guqq \lambda_{u}\etau(x)+
\gsqq \lambda_{s}\etas(x) ]\ q(x-y/2).
\label{S_Q_Pi_1}
\end{eqnarray}
The dynamic vertex $F\left[x+y/2,x-y/2;\Lambda_{\rm IQM} ^{-2}\right]$ depends on the coordinates of the quark and
antiquark and arises due to the quark-antiquark interaction induced
by exchange of instanton with
size $\Lambda_{\rm IQM}^{-1}\simeq\rho_c$, $\Lambda_{\rm IQM}\approx 0.742$ GeV.
The nonlocal vertex characterizes the coordinate dependence of order
parameter for SBCS
and can be expressed in
terms of the nonlocal quark condensates.

We restrict ourselves to the approximation
(see, e.g. \cite{Gross90})
\begin{equation}
F\left[x+y/2,x-y/2;\Lambda_{\rm IQM} ^{-2}\right] \to F(y^2, \Lambda_{\rm IQM}^{-2}),
\label{Separable}
\end{equation}
when the dynamic quark-meson vertex depends only on the relative coordinate
of the quark and antiquark squared, $y^2$, if neglecting the dependence of
the vertex on the angular variable $(yx)$. The Fourier transform of the vertex
function in the Minkowski space is defined as $\tilde F(k^2;\Lambda_{\rm IQM}^{2}) =
\int d^4x F(x^{2};\Lambda_{\rm IQM} ^{-2})\exp(-ikx)$ with normalization $\tilde F
(0;\Lambda_{\rm IQM}^{2}) = 1$, and we assume that it rapidly decreases in the
Euclidean region ($k^2=-k_E^2\equiv -u$).
As in the NJL model, we also approximate the
momentum-dependent quark self-energy in the quark propagator $S^{-1}(k;m_a)=%
\not\!k-m_a$ by a constant quark mass~\cite{Gross90} and neglect
meson mass effects. The quark masses are $m_u=275$ MeV, $m_s=430$ MeV,
close to those obtained in the NJL model (see Sect.~3.1.).
We have to note that the approximations used here are
not fully consistent.
Further, as the reader will see below, the choice of the model for the quark-meson vertex
(\ref{Separable}), depending only on the relative coordinate, induces a
certain artifact in the $x$ behaviour of DA.
However, these
deficiencies of the chosen approximation
are not essential for the present
purpose and do not lead to large numericas errors.

The quark-meson coupling is
given by the  condition~\cite{Gross90}
\begin{equation}
g_{q}^{-2}=\frac{N_{c}}{8\pi ^{2}}\int\limits_{0}^{\infty }\!\!du\,u\tilde F
^{2}(-u;\Lambda_{\rm IQM}^{2}/m_q^2)\frac{3+2u}{\left( 1+u\right) ^{3}};  \label{gpiqq}
\end{equation}
and the meson weak decay constants
$
f_{\pi}\equiv \fu, \quad\mbox{and}\quad \fs
$
are expressed by
\begin{equation}
f_{q}=\frac{N_{c}g_{q}}{4\pi ^{2}}M_{q}\int\limits_{0}^{\infty}\!\!du\,u
\tilde F(-u;\Lambda_{\rm IQM}^{2}/m_q^2)\frac{1}{\left( 1+u\right)^2}.  \label{f_pi}
\end{equation}
We have rescaled the integration variable by the quark mass squared.
Within the instanton vacuum model, the size of
nonlocality of the nonperturbative gluon field,
$\rho _{c}\sim \Lambda_{\rm IQM} ^{-1}\simeq0.3$ fm,
is much smaller than the quark Compton length $m_{q}^{-1}$.

To calculate the transition form factors, one can use the same formulas
that have been obtained above in the NJL model. The only exception
is that  the $\theta$-function should be replaced by the nonlocal
quark-meson vertex function $\tilde F(-u,\Lambda_{\rm IQM}^2/m_q^2)$.
In the next section, we discuss numerical results obtained in IQM
as compared with the NJL model calculation and experimental data.

Let us note that we use an approximation to the model with constant
constituent quark masses for all three quark lines in the diagrams of the
process (see Fig.~\ref{Pgg}). However, the asymptotic result (\ref{Igpigg2}) is independent of
the mass  in the quark propagator with hard momentum
flow, as it should be. The other two quark lines remain soft during the
process; thus, the mass  $m_{a}$ can be considered as given on a
certain characteristic soft scale in the momentum-dependent case $m_{a} $.

\section{Discussion and conclusion}

Within the two ECQM under consideration
that describe the quark-meson
dynamics, we calculated the $ \gamma^{\ast} \to P\gamma^{\ast} $ transition form
factor at moderately high momentum transfers squared in a wide
kinematic domain. From the model
calculations, the  normalization
coefficient $J_P(\omega)$ of the leading $Q^{-2}$ term is found (see
(\ref{Igpigg2})).
It depends on the ratio of the constituent
quark mass to the UV cut-off $\Lambda$ and also on the kinematics of the process.
From the comparison of the kinematic dependence of the
asymptotic coefficient of the transition  form factors, given by pQCD  and NJL,
the meson distribution amplitudes (\ref{WF_VF}) are derived.
Analogously, a relation between DA and the dynamic quark-meson vertex
function is obtained in IQM.
In the specific case of symmetric kinematics ($q_1^2=q_2^2$),
our result agrees with the one obtained by OPE and also with
the expression for the constant $f_P$ that determines the  decay $P\to\mu\tilde\nu_\mu$ of
meson $P$.

Let us discuss DA obtained in different approaches. In the NJL, we obtained
explicit expressions for the DA of pion, $\eta$, and $\eta'$ meson. They are
plotted in Figs.~\ref{fiuplot}--\ref{fietaplot} (solid line) from where
one can see that the NJL model predicts a more flat distribution of
quark momenta in a meson than IQM.
To  compare with other theoretical approaches, we also plotted, as a sample, the results
obtained in IQM (dashed line). There are also given the asymptotic pQCD
expression for  DA: $\phi^{\rm asympt}(x)=6x(1-x)$ (dashed-dotted line).
One can see that for the most of
$x$, the  DA's shapes are similar in different approaches.
The difference can be noticed near $x\approx 1/2$ and at the edges.
The cusp at $x=1/2$ and a very sharp decrease of DA near $x=0$ and $x=1$
are  artifacts closely related to the UV regularization
in the NJL model and to the shape of the nonlocal quark-meson vertex function
in IQM. However, these deficiencies turned out to be not crucial in our
calculations.

Now, we would like to make some
notes regarding the definition of $f_{\eta}$ and $f_{\eta'}$
used by authors of  \cite{CLEO}.
 In \cite{CLEO}, $f_P$ are obtained from the tabulated
data on the decays $P\to\gamma\gamma$, using
the low-energy limit of the process amplitude
\begin{equation}
{\mathcal{F}}_P(0,0,0)\simeq 1/(4\pi^2\tilde f_P)
\label{TQ0}
\end{equation}
(see (6) in \cite{CLEO}).
This works well for the pion, but the case of the $\eta$ and $\eta'$
mesons is rather different because of the singlet-octet mixing.
In the limit $Q^2\to\infty$, one should expect:
\begin{equation}
\lim_{Q^2\to\infty}Q^2 {\mathcal{F}}_P(q_1^2=-Q^2,q^2=0,p^2=0)=2 f_P,
\end{equation}
where $f_P$ are not the same as $\tilde f_P$ except for $f_\pi$ as
one can see from comparing (\ref{feta}) and (\ref{fetaprime}) with
(\ref{f0eta}) and (\ref{f0etaprime}).
 That is why
the CLEO fit noticeably disagrees
with the limit $2f_{\eta'}$ (see (5) in \cite{CLEO})
drawn in Fig.~23 \cite{CLEO}. Therefore,
it is not correct to use the equation (7) in \cite{CLEO}
to perform a fit. From our calculation, we see that,
taking into account the singlet-octet mixing, one can
avoid the discrepancy in the description of the
$\gamma^\ast\to P\gamma$ interaction both at small and large $Q^2$.

Now we compare our results for the case $\omega=1$
($\gamma^\ast\to P\gamma$) with
those given by the CLEO collaboration \cite{CLEO}
for large $Q^2$. We calculate the product of $Q^2$ and
transition form factor:
\begin{equation}
{\mathcal{F}}_P(\omega)=\lim_{Q^2\to\infty}
Q^2{\mathcal{F}}_P(q_1^2,q_2^2,p^2).
\end{equation}
Theoretically we have
\begin{equation}
{\mathcal{F}}_P(\omega)=J_P(\omega) f_P ,
\end{equation}
according to (\ref{AmplAsympt}).
In the NJL model, $J_\pi(1)=1.97$, $J_\eta(1)=2.04$, $J_{\eta'}(1)=1.9$.
Therefore,
${\cal F}_\pi(1)\approx 0.184$ GeV,
${\cal F}_\eta(1)\approx 0.193$ GeV,
and ${\cal F}_{\eta'}(1)\approx 0.256$ GeV.
The IQM predicts $J_\pi(1)=1.78$, $J_\eta(1)=1.83$, $J_{\eta'}(1)=1.73$ and
${\cal F}_\pi(1)\approx 0.16$ GeV,
${\cal F}_\eta(1)\approx 0.17$ GeV,
and ${\cal F}_{\eta'}(1)\approx 0.23$ GeV.

\begin{figure}
\caption{}
\includegraphics[angle=90,scale=0.7]{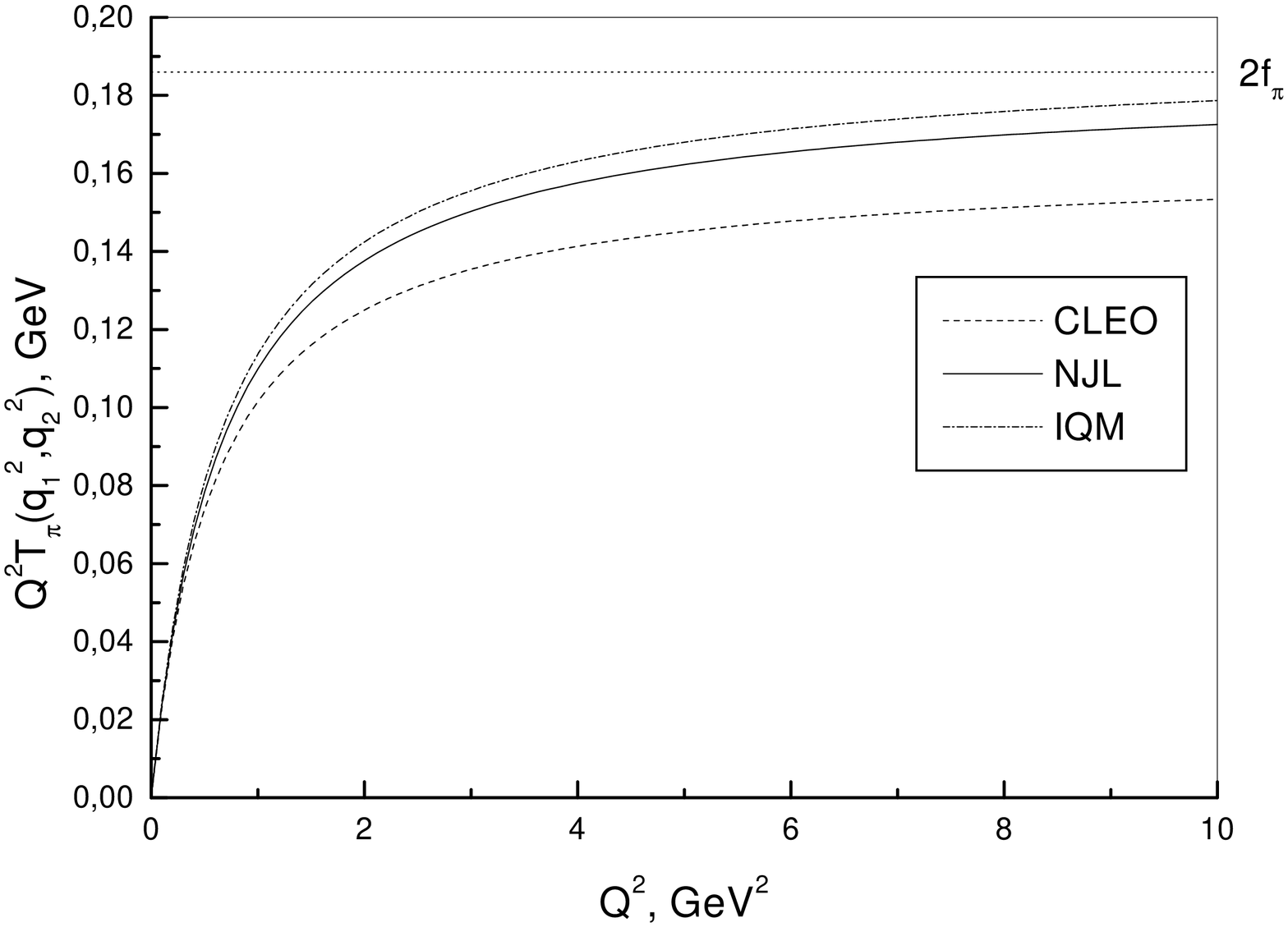}
\label{ffpi}
\end{figure}

\begin{figure}
\caption{}
\includegraphics[angle=90,scale=0.7]{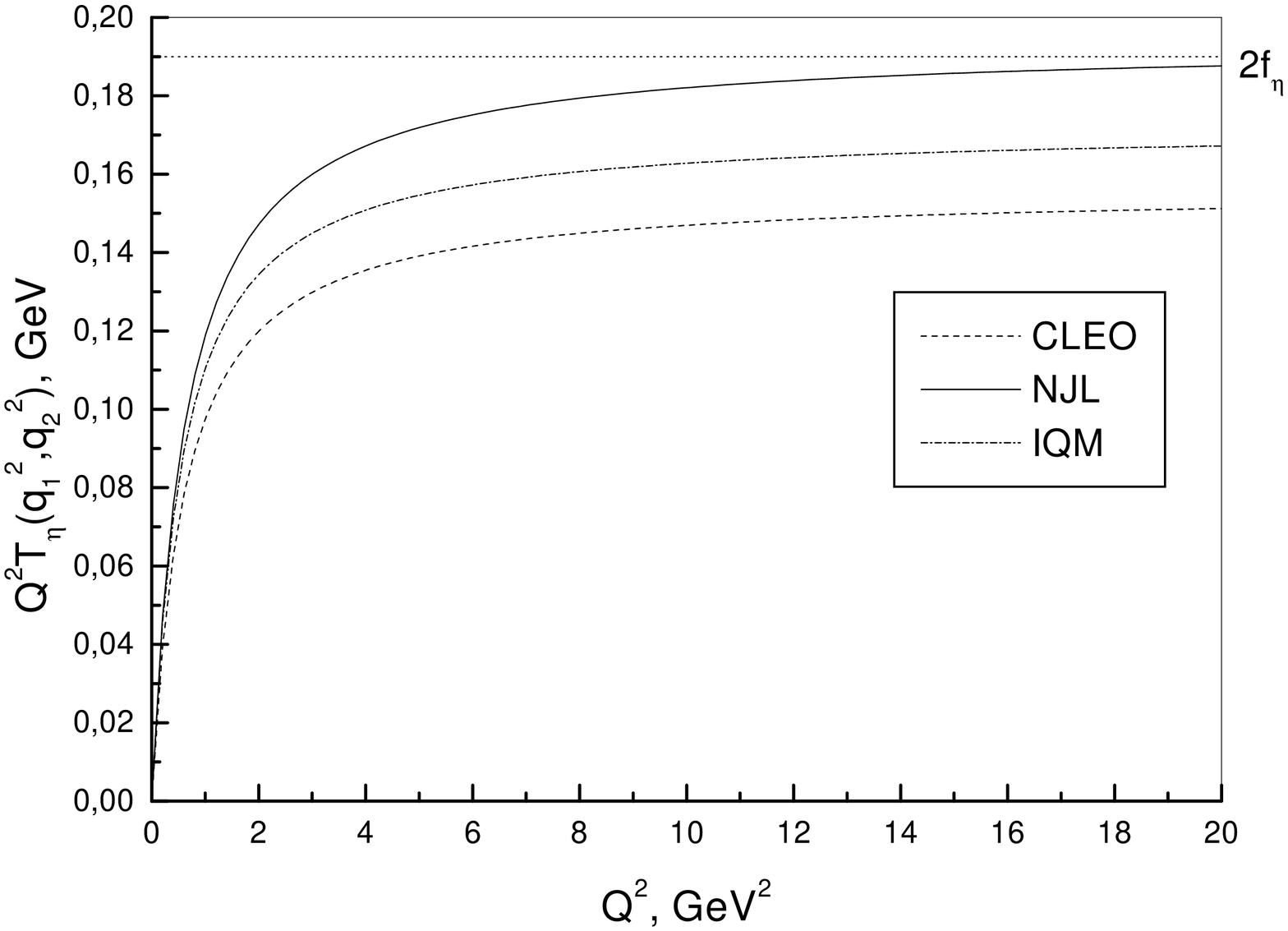}
\label{ffeta}
\end{figure}

\begin{figure}
\caption{}
\includegraphics[angle=90,scale=0.7]{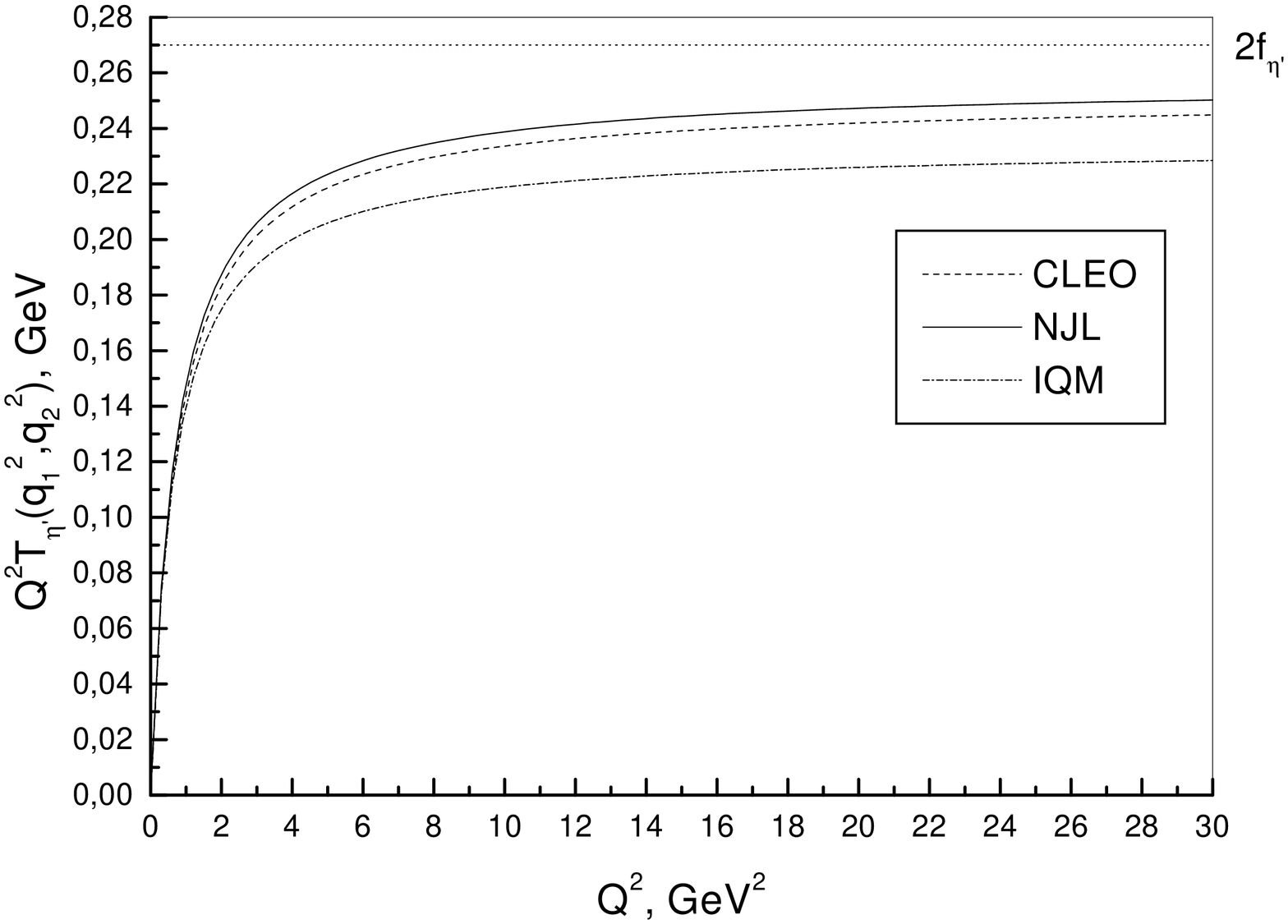}
\label{ffetaX}
\end{figure}

The infinite value of $Q^2$ cannot be reached in experiment,
so we determine ${\cal F}_P(\omega)$ at
 the value of $Q^2$,  maximum accessible in experiment.
Thus, the CLEO collaboration gives ${\cal F}_\pi(1)= 0.17\pm 0.3$ GeV
at $Q^2= 7.0-9.0$ GeV$^2$,
${\cal F}_\eta(1)\approx 0.16$ GeV at $Q^2\sim 22$ GeV$^2$,
${\cal F}_{\eta'}(1)\approx 0.25$ GeV at $Q^2\sim 30$ GeV$^2$.
The monopole interpolation for the transition form factors (\ref{Fpiggfit})
obtained both theoretically and  experimentally
is shown in Figs.~\ref{ffpi}--\ref{ffetaX}. One can see, that
the experiment gives  values for the transition form factors lower than
models. The biggest difference (about 20 \%) displays for the $\eta$ meson.
The best agreement of models with the CLEO fit is obtained for $\pi$ and $\eta'$.

The constants $g_{P\gamma\gamma}$ in the monopole Ansatz (\ref{Fpiggfit}) for
the transition form factors are related with the $P\to\gamma\gamma$ decay width.
It is interesting to compare their values predicted in different models
with experimental data. According to (\ref{TQ0}), we obtain from NJL:
$g_{\pi\gamma\gamma}=0.27$ GeV$^{-1}$,
$g_{\eta\gamma\gamma}=0.31$ GeV$^{-1}$, and
$g_{\eta'\gamma\gamma}=0.35$ GeV$^{-1}$.
The same gives us IQM.
From experiment we have
$g^{\rm exp}_{\pi\gamma\gamma}=0.26$ GeV$^{-1}$,
$g^{\rm exp}_{\eta\gamma\gamma}=0.26$ GeV$^{-1}$, and
$g^{\rm exp}_{\eta'\gamma\gamma}=0.34$ GeV$^{-1}$.
Again, the model prediction for $\pi$ and $\eta'$
better suits to the experimental values, whereas for
the $\eta$ meson, one has a noticeable discrepancy.

The  results presented in our paper are in
accordance with the conclusions made in \cite{MikhRad90,RadRus96,MikhBak98}
within the QCD sum rules, which evidences that
our approach is valid for the process under consideration and
makes us hope that it can be applied to other processes with
large momentum transfers, as well.

We plan to use the ECQM approach approved here on the process $\gamma^\ast\to P\gamma$,
for  the description of the following processes:
$\gamma^\ast\to P\rho$, $\gamma^\ast\to P\omega$, $\gamma^\ast\to P'\gamma$
(with $P'$ being a radial excitation of a pseudoscalar meson),
$\gamma^\ast\to S\gamma$ (with $S$ being a scalar meson),
$\gamma^\ast\to\gamma A$ (with $A$ being an axial-vector meson),
$\gamma^\ast\to\gamma\pi\sigma$, $\gamma^\ast\to\gamma\pi\rho$,
$\gamma^\ast\gamma^\ast\to\pi\pi$ etc.
\bigskip

The work is supported by the Heisenber-Landau program and RFBR Grants
01-02-16231 and 02-02-16194 and by Grant INTAS-2000-366.

\clearpage
\section*{Figure captions}
\begin{enumerate}
\item
The diagrams contributing to the process
$\gamma^\ast\gamma^\ast\to P$ ($\gamma^\ast\to P\gamma^\ast$) amplitude.
\item
The DA $\varphi_{\rm u}$ in the NJL and IQM models and
the asymptotic DA.
\item
The DA  $\varphi_{\rm s}$ in the NJL and IQM models
and the asymptotic DA.
\item
The DA  $\varphi_{\eta}$ in the NJL and IQM models
and the asymptotic DA.
\item
The DA  $\varphi_{\eta'}$ in the NJL and IQM models
and the asymptotic DA.
\item
The light-cone transition form factor for the pion.
The solid line corresponds to the NJl-model calculation,
the dashed line is a fit to the CLEO data, the dashed-dotted line is corresponds to
IQM, and the dotted
line is $2f_\pi$.
\item
The light-cone transition form factor for the $\eta$ meson.
The solid line corresponds to the NJl-model calculation,
the dashed line is a fit to the CLEO data, the dashed-dotted line is corresponds to
IQM, and the dotted
line is $2f_\eta$.
\item
The light-cone transition form factor for the $\eta'$ meson.
The solid line corresponds to the NJl-model calculation,
the dashed line is a fit to the CLEO data, the dashed-dotted line is corresponds to
IQM, and the dotted
line is $2f_{\eta'}$.
\end{enumerate}

\clearpage
\section*{Figures}


\end{document}